# Experimental Investigation of a Recurrent Optical Spectrum Slicing Receiver for Intensity Modulation/Direct Detection systems using Programmable Photonics

Kostas Sozos, Francesco Da Ros, *senior member, IEEE, senior member, Optica*, Metodi P. Yankov, *senior member, IEEE, member, Optica*, George Sarantoglou, Stavros Deligiannidis, Charis Mesaritakis and Adonis Bogris, *senior member Optica*
(Invited paper)

*Abstract*— **Photonic computing and signal processing are rapidly regaining attention, capitalizing on the saturation of digital electronics capabilities, owed to the ending of the Moore's law era. Optical communications are among the fields in which digital signal processing (DSP) struggles to offer a viable long-term solution, as its power consumption rises significantly in the latest generation optical links. For example, the traditional intensity modulation and direct detection (IM/DD) systems suffer from challenging problems as the chromatic dispersion induced power fading effect, which necessitates the use of powerful digital equalizers in the 800G and 1.6T optical transceivers.**

**In this paper, we experimentally validate our previous numerical works in recurrent optical spectrum slicing (ROSS) accelerators for dispersion compensation in high-speed IM/DD links. For this, we utilize recurrent filters implemented both through a waveshaper and by exploiting novel integrated programmable photonic platforms. Different recurrent configurations are tested. The ROSS accelerators exploit frequency processing through recurrent optical filter nodes in order to mitigate the power fading effect, which hinders the transmission distance and baudrate scalability of IM/DD systems. By equalizing even 80 km of 64 Gb/s PAM-4 transmission in C-band, we prove that our system can offer an appealing solution in highly dispersive channels. We employ the simplest digital equalization in the form of a feed-forward equalizer (FFE), avoiding throughput, latency, and complexity restrictions that the decision feedback equalizer (DFE) and the maximum-likelihood sequence estimator (MLSE) impose. We achieve, with the use of programmable photonics, to reduce the bit error rate (BER) from 0.11 to less than 1x10$^{-2}$ with the use of only two filter nodes, whereas BER approaches 10$^{-3}$ when three nodes are incorporated. These results correspond to almost two orders of magnitude BER gain. © 2024 The Authors**

## I. INTRODUCTION

NEUROMORPHIC photonics has emerged as an attractive way to overcome the inherent barriers of Von Neumann computing, by exploiting established concepts from traditional deep learning and combining them with the unrivaled attributes of light [1], [2], [3], [4], [5], [6]. Optical communications constitute an ideal candidate for the wider adoption of photonic information processing, as digital signal processing (DSP), which garners an ever-increasing role in next generation transmission systems [7], is accompanied by a vast increase in power consumption. For example, the present generation 800G IM/DD systems operate with 8 lanes of 50 Gbaud PAM-4 for up to 10 km in O-band [8]. The transition to 100 Gbaud lanes will be probably accompanied by 1-2 tap maximum likelihood sequence estimators (MLSEs) [9], [10], which, although minimalistic, roughly doubles the complexity and the power consumption of the DSP module [11]. Neuromorphic propositions that offer unconventional signal processing in order to combat nonlinearities or nonlinear dispersion in optical communications have been already presented in the literature [12], [13], [14], [15]. Although they all contribute to the field and clearly show the potential of photonic neuromorphic processing in compensating for dispersion and nonlinearity, most of the works do not substantially address how the proposed solutions promote low-power consumption, compatibility to very high baud rates, fabrication friendly and compact implementations in order to

This work was funded has been supported by the Hellenic Foundation for Research and Innovation (H.F.R.I.) under the "2nd Call for H.F.R.I. Research Projects to support Faculty Members & Researchers" (Project Number: 2901). This work has been also supported by the EU Horizon Europe PROMETHEUS project (101070195) and the Villum foundation (OPTIC-AI grant n. VIL29344) *(Corresponding author: A. Bogris).*

K. Sozos, S. Deligiannidis and A. Bogris are with the University of West Attica, Dept. of Informatics and Computer Engineering. Aghiou Spiridonos, 12243, Egaleo, Athens, Greece, ksozos@uniwa.gr, abogris@uniwa.gr

C. Mesaritakis and G. Sarantoglou are with the University of the Aegean, Dept. of Information and Communication Systems Engineering, Palama 2, Karlovassi, 83200-Samos-Greece, cmesar@aegean.gr

F. Da Ros and M.P. Yankov are with the Technical University of Denmark, Department of Electrical and Photonics Engineering, Ørsteds Plads Building 343, DK-2800 Kgs. Lyngby, Denmark, {fdro, meya}@dtu.dk

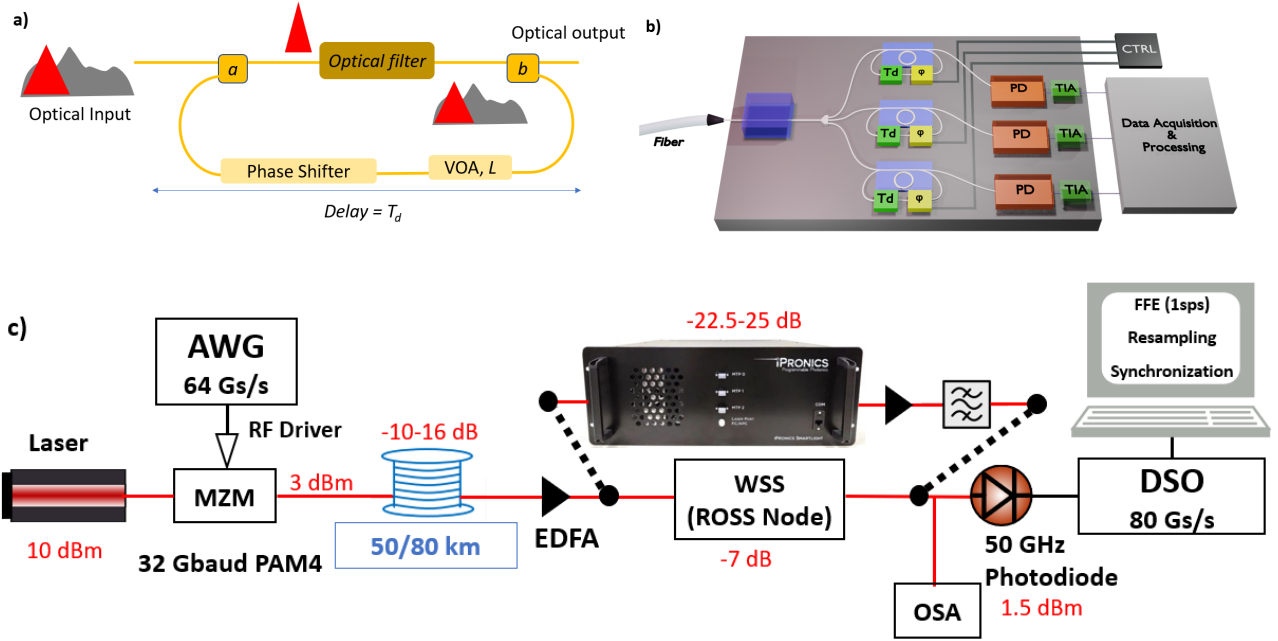

Figure 1. a) The recurrent filter node that constitutes the main component of ROSS-NN. b) A three-node ROSS receiver that can be used for both IM/DD and self-coherent configurations. c) The 32 Gbaud experimental setup, with the ROSS nodes implemented through a WSS or with the use of iPronics Smartlight. In the second case, an extra EDFA and an optical filter are utilized in order to compensate for the higher insertion losses.

have their chance at offloading DSP in next generation short-reach transmission systems.

In order to bridge the gap between practical implementations and neuromorphic photonic processing, very recently, we have proposed recurrent optical spectrum slicing neural networks (ROSS-NN), that can be tailored to address important issues of high-speed communications (dispersion-induced power fading effect, nonlinearity mitigation) with a small number of optical components (optical filter nodes) and potentially minimal power consumption, thus allowing for a smooth adoption in pluggable transceiver modules [16], [17]. Although this technique has shown its potential in numerical simulations for both intensity modulation – direct detection (IM/DD) and self-coherent systems operating at 200 Gbaud, there has been no experimental verification of the results to date due to the lack of commercially available reconfigurable recurrent filters. Reconfigurable photonic platforms [18], [19] that can support very high bandwidths meeting the needs of 800 G and beyond era, provide great implementation flexibility for the testing of various promising photonic schemes, reducing the time between conception and realization. So, they can contribute to the adoption of neuromorphic ideas in multiple fields especially if custom designs will be adopted in order to reduce excess losses [19].

In this work, we present an experimental investigation of our previously proposed ROSS receivers in IM/DD systems. More specifically, we employ different recurrent structures, either implemented by a programmable wavelength selective switch (WSS) or with the use of a reconfigurable photonic platform (iPRONICS Smartlight product) and use them in order to equalize a vastly distorted 32 Gbaud PAM-4 signal after 50-80 km of C-band transmission. In section II, we briefly present the concept of operation of ROSS receivers and the experimental testbed. In section III, we provide the experimental results for the equalization of the 64 Gb/s PAM-4 signal. We present separately the results with the selected filters emulated by the WSS and the ones generated with the use of programmable photonic circuits with two different recurrent configurations, as well as the comparison between the two different types of filter nodes. Finally, we conclude the work in section IV.

## II. Conceptual Presentation and Experimental Setup

### A. ROSS receiver concept for IM/DD links

The main idea behind ROSS-NN lies in processing signals which exhibit rich spectral content directly in the frequency domain. In these cases, it makes sense to use neural nodes that can separate and manipulate the frequency information, something that is also present in the biological neurons [20]. This concept can degenerate to simpler neuromorphic accelerators which can process telecom signals in order to amend frequency selective impairments. The frequency-selective (colored) noise is the major practical limitation in most high-speed IM/DD systems. It stems from the limited component bandwidth, which attenuates the higher signal frequencies, or from the dispersion-induced power fading [21], [22]. In this second case, the periodicity of the signal attenuation creates a significant barrier that even the stronger digital equalizers cannot fully overcome [23]. Our previous works have numerically showcased improved performance in dispersive IM/DD links, by the usage of one or two node ROSS receivers [11], [16], whilst even simple filtering proposed in [24], [25] has shown that power fading can be partially handled. In [16], the improvement was attributed to the modular transfer function of the recurrent filters, which can provide precise attenuation or enhancement of different frequencies, while also providing time-dependencies determined by the feedback loop, which controls the fading memory of the filter node.

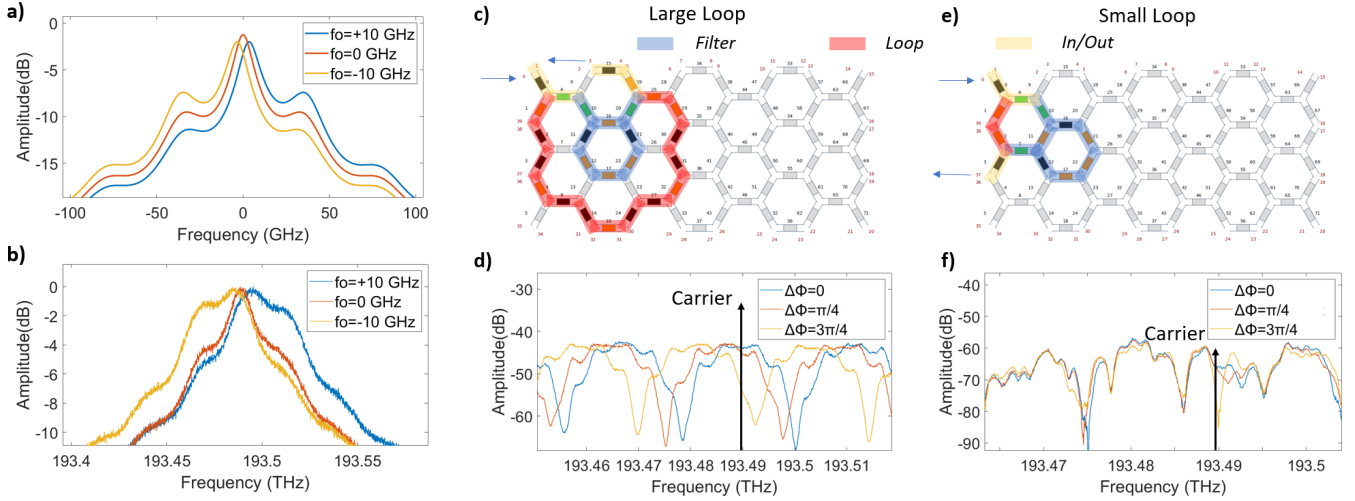

Figure 2. The analytically calculated provided to the WSS a) and the generated transfer functions in an optical spectrum analyzer b). c). e) The two configurations of ROSS nodes in iPRONICS Smartlight and the corresponding measured transfer functions d), f).

The recurrent filter node is depicted in Fig. 1a, while its versatile transfer function can be described by the Eq. (1).

$$H_{node}(f) = \frac{\sqrt{1-a}\,\sqrt{1-b}\;H(f)}{1+\sqrt{a\,b\,L}\;H(f)\;e^{-i(2\pi f\,T_d+\varphi)}} \qquad (1)$$

with $H(f)$ being the transfer function of the filter without any feedback presented in Fig. 1a. In this work we implement $H(f)$ as a recurrent $1^{st}$ order Butterworth filter using the programmable WSS and as a recurrent Mach-Zehnder delayed interferometer (MZDI) using the programmable photonic platform. The transfer functions of the two basic filter units are provided in Eq. (2), (3) for the Butterworth and the MZDI respectively.

$$H(f) = (1 + i\;e^{-i(2\pi(f-f_0)\Delta T)})^{-1} \qquad (2)$$

$$H(f) = \frac{1 + e^{-i(2\pi(f-f_0)\Delta T)}}{2} \qquad (3)$$

In Eq. (1), the parameters $a$ and $b$ are the coupling strengths of the input and output coupler. In Eq. (1), (2), $1/\Delta T$ is the cutoff frequency of the filter. In Eq. (3), $\Delta T$ is the time delay between the two arms of a MZDI. The parameter $f_0$ is related to the frequency shift with respect to the central frequency of the signal. $L$ refers to the feedback strength, i.e the portion of the filter output that is fed back in the input. In a passive configuration such as the one we employ here, $L<1$. Lastly, the $T_d$, and $\varphi$ are the time delay and the phase shift characterizing the feedback loop. Versatile spectral processing can be achieved by tuning the $\varphi$ and $f_0$ and $L$ parameters of this node, as shown in [16]. With the use of reconfigurable photonics, $\Delta T$ and $T_d$ can be also varied to some extent. In Fig. 1b, an indicative three-node receiver that can be used for reception and equalization of IM/DD or self-coherent systems is illustrated. An optional integrated optical amplifier can be placed before the splitter in order to compensate for the insertion and splitting losses.

### B. Experimental setup

The experimental setup that is used in this work, is depicted in Fig. 1c. At the transmitter side, a pseudorandom unrepeated PAM-4 sequence is produced in Matlab for two different random seeds. The PAM-4 sequences are sequentially uploaded to an arbitrary waveform generator (AWG) with 23 GHz analog bandwidth, at 2 samples per symbol (sps). The resulted signal is resampled to the sampling rate of AWG, operating at 64 Gsa/s. A 55 GHz radio frequency (RF) driver amplifies the electrical signal and a Mach-Zehnder modulator (MZM), with analog bandwidth of 25 GHz is used to modulate the signal with the bias set at the quadrature point. The modulated carrier is produced by a tunable C-band source operating at 1549.4 nm, with 10 dBm of output power. The optical signal, with 3 dBm launched power after the MZM, is propagated through different lengths of single-mode fiber (SMF) and then is amplified by an erbium-doped fiber amplifier (EDFA). In this work, we study transmission at 50 km and 80 km spools, with dispersion coefficient D=16.4 ps/nm/km.

The ROSS nodes follow after the amplification, implemented either with the use of a WSS, or with a programmable photonic platform (iPronics Smartlight processor). In the first case, the applied transfer functions are produced through Eq. (1), (2) with $\Delta T$=50 ps and $T_d$=21.875 ps, while $f_0$ ranges from +/-1 to +/-15 GHz. The provided and the produced transfer functions are depicted in Fig. 2a,b. In the second case, the reconfigurability of the platform through programmable unit cells (PUCs), permits the implementation of two node configurations, incorporating a MZDI with 4 PUCs path difference ($\Delta T$=45 ps) as the optical filter and two different feedback loops. This path difference determines the free spectral range (FSR) of the filter and the subsequent bandwidth, which are 22.2 GHz and 11.1 GHz respectively. The first loop employs 14 PUCs which correspond to ~11.36 mm feedback length and 157.5 ps delay, while the second one is constructed from only 2 PUCs, with length 1.622 mm and delay 22.5 ps. The two different configurations and their corresponding spectral responses are depicted in Fig. 2c,e and 2d,f, respectively. The two loop configurations will be called large and small loops, respectively in the rest of the manuscript. Each of the two configurations is also implemented as a simple interconnect, namely as a direct, non-recurrent connection between input and output, for comparison. The small loop interconnect includes 4 PUCs while the large one, 7 PUCs. The WSS (Waveshaper 4000s) reduces the out-of-band noise significantly and exhibits 7 dB losses when implementing the ROSS node with $f_0$=0. On the other hand, the overall Smartlight losses are 25 dB for the large and 22.5 dB for the small loop, when the phase shift of the filter is zero, necessitating the use of a second EDFA followed by an extra 1-nm wide optical filter to remove out-of-band noise. Then, a 99-1 coupler feeds an optical spectrum analyzer (OSA, low-power tap) and a photodetector of 50 GHz bandwidth in both the WSS and Smartlight cases. The received optical power ranges from approximately -0.5 to 4 dBm, depending on the use of WSS or Smartlight, and the frequency offset of the filter with respect to the carrier. Lastly, an 80 GSa/s digital signal oscilloscope (DSO) is utilized, capturing 3 instances of the received data series for each transmitted seed, and for each filter under test which are sequentially measured. The receiver-side offline processing includes resampling to 2 sps, time synchronization carried out separately for each filter position and downsampling to 1 sps in most of the cases as we want to focus on the most practical (low complexity) and cost-efficient symbol processing approach. The selected synchronized and resampled outputs are provided as inputs to a symbol-spaced FFE with 25 taps. The number of taps was chosen after an optimization process showing performance saturation for a larger number of taps. We launch up to 3 filter outputs in the FFE post-processor therefore the final weights scale from 25 (1 filter output) to 75 (3 filter outputs). We use 10000 symbols for training via a linear regression algorithm. PAM-4 demodulation follows and BER is calculated by direct error counting. We use 45000 symbols (90000 bits) for testing and we average the BER for the 3 DSO captures of each seed.

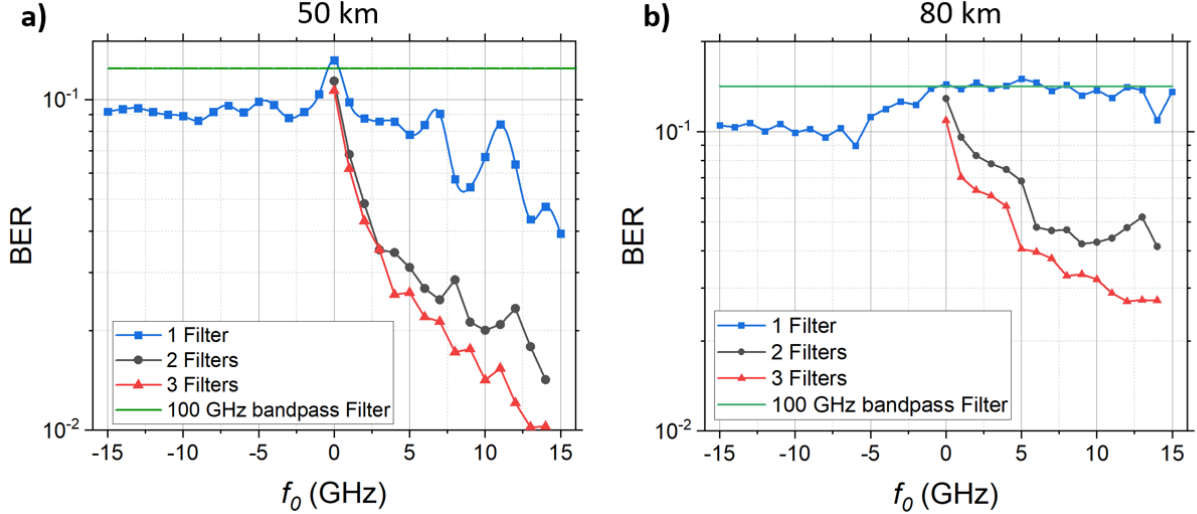

Figure 3. WSS results for the two tested transmission distances. In each case, the single filter results are shown along with the two and three node results. The all-pass filter serves as a fair benchmark for the BER improvement.

## III. EXPERIMENTAL RESULTS OF THE ROSS BASED RECEIVER IN A 32 GBAUD PAM-4 LINK

### A. ROSS nodes implemented through a WSS

At first, we examine the performance of the ROSS nodes implemented through the WSS. For this task, we employ 31 transfer functions for $f_0$ values between -15 GHz and 15 GHz with 1 GHz step. In Fig. 3, we provide BER results for the two examined transmission distances. Here, the single filter results provide only minor performance improvements, for $|f_0|$ values >5 GHz. The two-filter results are presented as a function of their symmetrical offset from the $f_0$=0 filter following the approach presented in [16], which dictated that the symmetrical positioning of the filters with respect to the signal carrier frequency provides excellent performance when a Butterworth in a loop is considered. When we use a third filter, we position it at $f_0$=0 GHz, i.e. centered at the signal carrier. This choice is related to the improvement of the signal to noise ratio (SNR) that is provided by adding the third filter, especially when the two other filters are largely detuned from the carrier. We can notice the expected significant performance improvement when the detuning of the two filters increases, providing higher diversity in the spectral domain. The addition of a third filter, further enhances the performance only when the two detuned filters are well separated (frequency offset > 10 GHz), because for small offsets there is significant overlap in the information that each filter provides to the receiver and for large offsets the SNR is deteriorated without the addition of the third filter. BER-wise, the results for the 100 GHz bandpass filter with the FFE are above $10^{-1}$ even for 50 km distance, attributed to the strong power fading effect. For the 50 km case, the best that a single detuned node can provide is $4.2 \times 10^{-2}$ while the BER drops below $2 \times 10^{-2}$, which is the frequently used soft-decision open forward error correction (O FEC) limit [26], with two filter nodes. With two nodes, BER equals to $4 \times 10^{-2}$ at 80 km, while the addition of a third node further improves the BER to $2.7 \times 10^{-2}$. In all these cases, the proposed two-node ROSS receiver improves the results by an order of magnitude, with respect to the linear digital equalization, however the improvement does not reach the full gain predicted in the numerical simulations presented in [16], for reasons that will be explained in the section III.C.

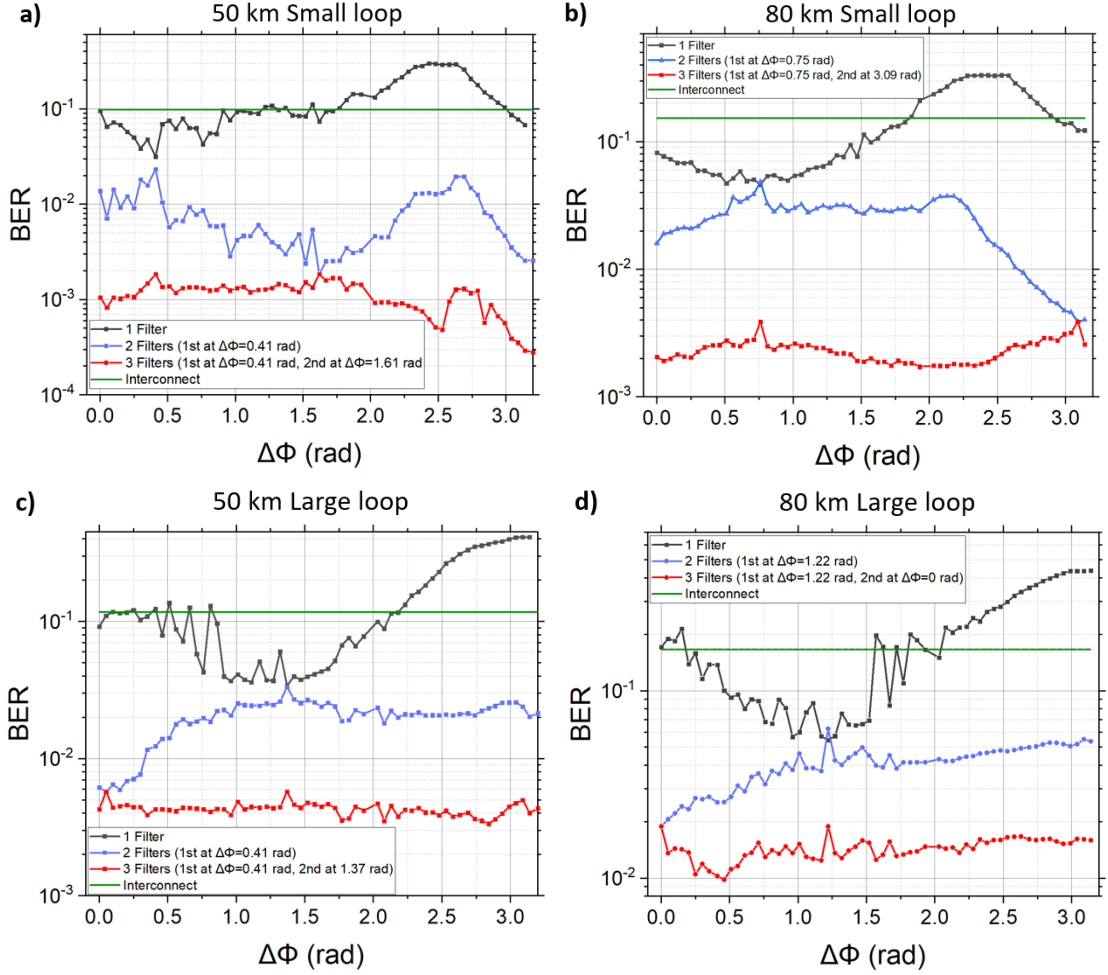

Figure 4. Experimental results with recurrent nodes implemented through the Smartlight processor. a-b) BER performance of the small loop configuration for 1-3 filters in the two different transmission distances as a function of the imposed phase shift ΔΦ in the MZDI filter. In the two and three node cases, this ΔΦ is chosen in respect of a central ΔΦ=2.2 rad. c-d) The comparison between the two different architectures of recurrent filter nodes. The node with the smaller feedback loop performs systematically better for large phase shifts.

### B. ROSS nodes implemented with the use of a programmable photonic platform

In this subsection, we examine the performance of the same ROSS receiver, with one, two, or three recurrent nodes, with the use of the programmable photonic processor. As mentioned in Section II, we chose to implement an MZDI filter, employing two paths of one and five PUCs respectively and we construct two different architectures, with different loop delays. The only parameter that is changed in this work to tune $f_0$ is the phase difference of the two paths in the MZDI filter, $\Delta\Phi$. The results are presented with respect to this phase shift between the branches of the MZDI filter, which has an equivalent effect to the parameter $f_0$ of the previous section. In the study with the WSS, we followed the approach of the antisymmetric positioning of the band pass filter nodes generated by the WSS with respect to the signal carrier in order to process separately the lower and the upper sideband of the signal. In the case of the MZDI recurrent filters, the nodes exhibit periodic transfer functions as depicted in fig. 2d,f. Therefore, unlike for the WSS case, we chose a simple optimization process which progressively selects the proper filter nodes in an iterative manner in all cases. Initially, we scan all the filter nodes (62 in total, with a ΔΦ step of 0.05 rad) and select the single filter offering the best BER performance at each transmission distance for the small and the large loop configurations. For the two-filter case, we combine this specific filter with all the remaining 61 positions and select the second filter which maximizes the overall BER gain. Similarly, in the three-filter receiver case, we keep the best two filters from the previous iteration process and scan exhaustively the remaining 60 positions in order to find the best three-filter combination.

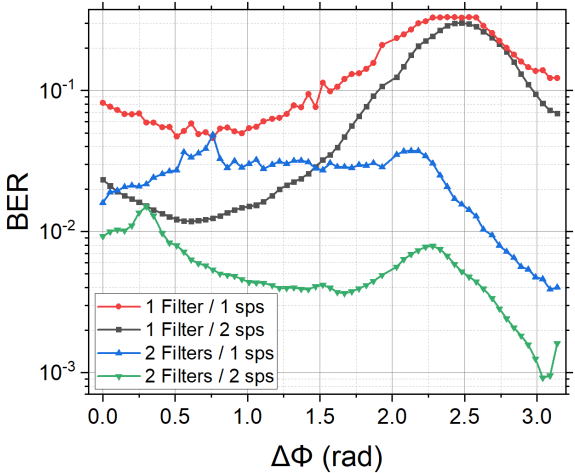

Figure 5. BER performance for one and two sps post-processing. These results refer to the Smartlight implemented filter with the small feedback loop at 80 km transmission. FFE with two sps improves BER by an order of magnitude.

In Fig. 4, we provide the results for the small loop (a, b) and for the large loop configuration (c, d), at 50 km and 80 km transmission. For comparison, the performance of the system when the Smartlight is in the interconnect configuration where no filtering takes place (aside from the bandpass filter before the PD) is also shown. Here a single filter can improve BER by more than 3 times, approaching $3 \times 10^{-2}$ at 50 km and $4.7 \times 10^{-2}$ at 80 km transmission. The two-filter receiver evidently improves the performance by over an order of magnitude with respect to the single detuned filter, while the addition of a third filter gives a smaller improvement in the order of 2-3 times. It is worth mentioning that for the small loop case, at 50 km, the third filter improves the BER an order of magnitude compared to the two-filter case. At 80 km, the two-filter receiver achieves BER=$3.9 \times 10^{-3}$ and the three-filter one BER=$1.7 \times 10^{-3}$, near and well below the hard-decision FEC limit of $3.8 \times 10^{-3}$ [10], respectively, which is a significant result showing the potential of ROSS in treating dispersion effects efficiently. In all these cases, the best performance is achieved when the second filter is sufficiently detuned and thus decorrelated from the first, showing that the performance improvement when combining filter outputs at the FFE algorithm is attributed to the frequency diversity and not just in the SNR improvement. The same also stands for the third filter. In order to confirm this assumption, we average our three DSO captures for the interconnect case and find that the BER improvement does not surpass 5%. So, in the interconnect case, the system is clearly dispersion limited. On the other hand, when the power fading is mitigated by the frequency processing, the averaging of the 3 DSO captures improves performance by 40%, showing that the ROSS receivers in this experimental work are mostly SNR limited, most probably by the high losses introduced unavoidably by the programmable photonic processor and the subsequent additional amplification. In Fig. 5, we present a comparison of the BER performance with 1sps and 2 sps processing for 80 km distance, showing that the ROSS receiver can even further boost BER performance when more than one samples per symbol are directed to the linear equalizer. In these cases, the performance improves by almost 5 times, both for the single filter and the two-filter accelerators.

TABLE I

| Two Filter Receiver | Minimum BER |
| --- | --- |
| **1 sps WSS** | $4.2 \times 10^{-2}$ |
| **1 sps Smartlight Small** | $7.1 \times 10^{-3}$ |
| **1 sps Smartlight Large** | $5.2 \times 10^{-2}$ |
| **2 sps WSS** | $5.9 \times 10^{-3}$ |
| **2 sps Smartlight Small** | $9 \times 10^{-4}$ |
| **2 sps Smartlight Large** | $8.7 \times 10^{-3}$ |

### C. Comparison of the two methods and discussion

As it has already become apparent, there is a mismatch in the performance between the WSS and the Smartlight implemented nodes. Furthermore, the Smartlight nodes with the larger feedback loop underperform, in comparison with the small loop counterparts. Table I, presents this performance divergence in the case of 80 km transmission with two filters and one or two sps, while one can trace back to the Figs. 3 and 4 for further comparisons. The main reason behind this is the limited resolution of the waveshaper, that cannot reproduce the sharp secondary peaks in the transfer function of Eq. (1) which correspond to the feedback loop that transforms the filter to a recurrent node. The same is true for the Smartlight node with the large feedback loop, that increases the feedback losses through the 14 PUC roundtrip, attenuating the feedback signal and the sharp feedback peaks. In Fig. 6a, we plot the transfer characteristics of a recurrent filter, for different values of the feedback parameter $L$, that we call feedback strength. The higher this value, the sharper the frequency selectivity of the filter. We can assume that both the spectral responses of the WSS and the large loop Smartlight implemented filters, as shown in Fig. 2b,d, resemble to that of a node with very low feedback strength, close to $L$=0.1. For example, the large loop imposes losses over 0.5 dB per PUC [19], corresponding to at least 7 dB overall feedback losses. In order to verify this fact, we simulate our 80 km transmission link, with careful choice of the loss budget and the total amplification noise, as well as with the exact same functions that we load to the WSS. Then, we compare the results with the experimental ones, for the two-node receiver, in Fig. 6b. It is evident that increasing the feedback strength vastly improves the performance, justifying the low performance of the low-resolution generated filters using the WSS . On the other hand, the actual optical filters with the small loop, implemented with the use of the reconfigurable platform, exhibit higher extinction ratio. So, it is reasonable that their performance approaches the performance of the simulated system. It must be stressed out, that the $4 \times 10^{-3}$ BER plateau in the simulations, is owed to the high simulated losses. The proposed system performs error free, even for 80 km of transmission, for links with better SNR. However, the performance in this experiment is limited by the ~40 dB combined losses of the Smartlight processor and the transmission, for the case of the Smartlight implemented filters.

The lower capability of the WSS produced filters to mimic strong feedback loops required for recurrent processing in the spectral domain can be also concluded by the electrical power spectra of the received signal after the photodetection. As the ROSS receiver exploits spectral processing in order to decolor the noise before the FFE, the combined electrical spectral response of the signal should be clear from deep notches. However, in the electrical spectra of the signal after the WSS filters and the photodetection in Fig. 6c, the power improvement of the first four notches does not surpass 2.5 dB, from the initial signal (the one after transmission bandpass filter and photodetection). On the other hand, the Smartlight optical filters provide approximately 10 dB gain for the first three notches (Fig. 6d), due to their stronger feedback characteristics which can better decorrelate the transfer characteristics of the different nodes and thus the frequency where fading appears in each case. This decorrelation of the filter nodes in the frequency occurrence of the fading dip justifies the improved performance.

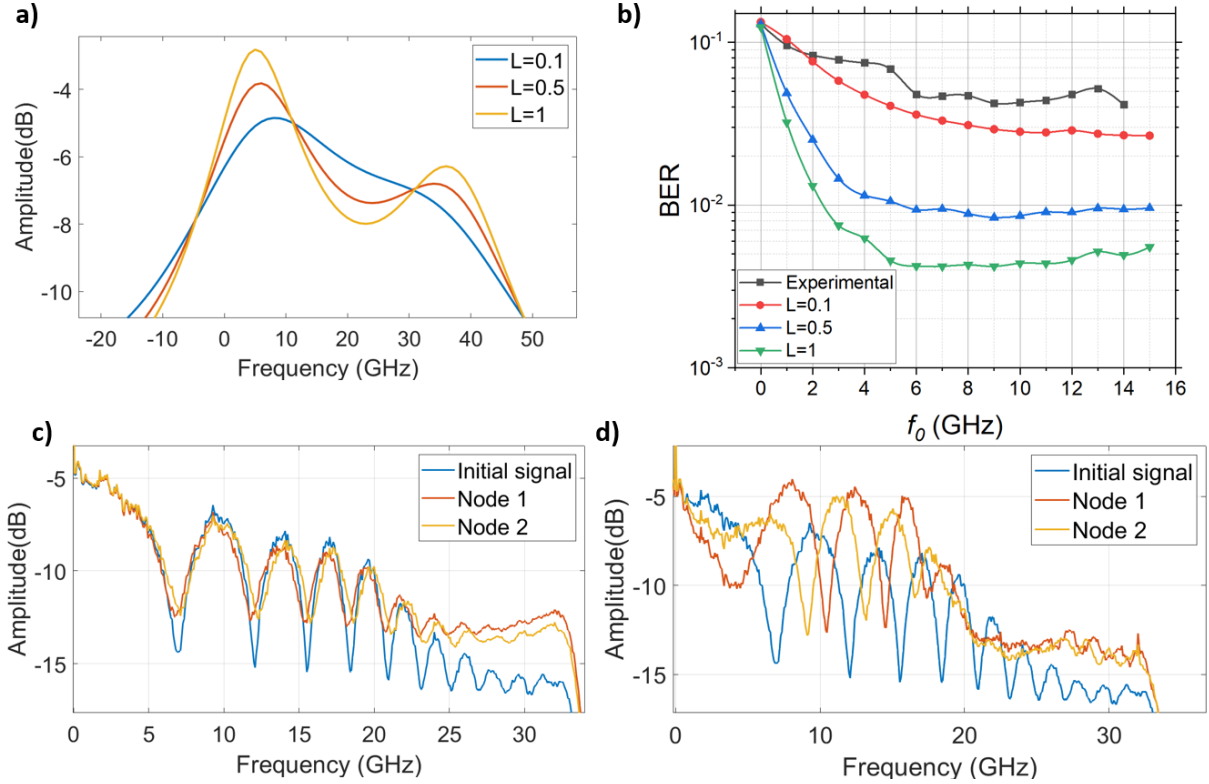

Figure 6. Comparison between the WSS and the Smartlight implemented filters. a) The transfer function of the ROSS node for different feedback parameter values and b) the subsequent simulated BER results for each case. The experimental results are given along with the simulation ones, in order to showcase the poor WSS resolution and its effect on the performance. c, d) The electrical spectra of the initial transmitted signal (after the 100 GHz bandpass filter) with and without the ROSS nodes for the WSS and the Smartlight cases respectively.

## IV. Conclusion

In this work, we provide, for the first time, experimental results for the equalization of a highly dispersive IM/DD link, with our previously proposed ROSS receiver. We employ recurrent optical filter nodes, emulated through a WSS and a programmable photonic platform and we showcase twentyfold performance improvement with the use of two recurrent nodes, utilizing a simple (2 nodes)×25 tap linear equalizer. At the best case, that of the small loop iPronics configuration, we achieve BER=$3.9\times10^{-3}$ at 80 km transmission, which is a pre-FEC BER compatible with HD-FEC codes. Further improvement can be accomplished with the aid of a third node or with stronger DSP, as with 2 sps equalization, achieving even BER= $9\times10^{-4}$ at 80 km of transmission in the C-band. The limiting factors in this experiment are two. The first limitation is the weak feedback of the recurrent nodes, due to the limited resolution of the WSS or the long and lossy feedback loop of the large-loop Smartlight configuration. The second limitation is the excess losses of a programmable processor which is the price one has to pay for the sake of reconfigurability and is vastly deteriorating the SNR performance. Nevertheless, this experiment clearly shows that a custom defined photonic chip implementing recurrent filtering behaviour with minimal losses has the potential to support the post-processing of high baud rate systems (>100 Gbaud) at low consumption and low latency.